# Orbit response measurements with oscillating trajectories

A. Shemyakin, Fermilab, Batavia, IL 60510, USA
*shemyakin@fnal.gov*

*Abstract*
Optical properties of components in a linac can be reconstructed from analysis of the orbit response matrix, composed from responses of BPMs to changes in currents of dipole correctors. An efficient way to record such response, discussed in this note, is to excite the corrector current with a sine wave and record its Fourier component of the BPM spectrum. Because of a narrow band of the excitation signal, response to multiple correctors can be recorded at the same time without significantly affecting the signal-to-noise ratio. For that, the correctors are oscillated at different frequencies, and the corresponding BPM Fourier components are recorded. In an ion (non-relativistic) linac, similar measurements can be performed for the longitudinal plane as well. Examples of measurements at the PIP2IT test linac are shown.

## Contents

## 1. Introduction

Optical properties of circular accelerators are routinely measured by the method of orbit response matrix [1]. In the simplest realization, it consists of recording the beam trajectory with Beam Position Monitors (BPMs) at two settings of a dipole corrector and comparing the difference between these two measurements ("orbit response") with prediction of the optical model. At Fermilab, the method is also referred as "differential trajectories" or "differential orbits" [2], and these terms are used interchangeably through this note. To reveal possible deviations as magnet calibrations, rolls etc. with a high accuracy, a "response orbit matrix" of responses of all BPMs to all correctors is recorded and compared with simulations. The most known procedure is LOCO (Linear Optics from Closed Orbits) [3] and its later modifications (see the review [4]).

Similar efforts appear to be less common in the linacs; one of the recent examples is optics measurements at LCLS [5]. Such measurements are particularly important at the initial commissioning or if unexpected problems with linear optics appear, allowing to find errors of wiring, calibration, or representation of the beam optics in the simulation model. For the same purpose, they were used at PIP2IT test accelerator [6] at various stages of its operation.

To provide low relative measurement error, the displacement induced by the change in the corrector current should be large enough in comparison with the BPM noise and beam jitter but may be limited by the available aperture and by preference to stay in the linear area of the magnets. One of the ways to counteract the noise, discussed in this paper, is to replace a single large displacement (or several displacements) by oscillating the corrector current (and therefore, the trajectory) with a sine wave of a small amplitude. Since the excitation is performed in a narrow band, it allows also to make measurements simultaneously with multiple correctors by oscillating them at different frequencies at the same time. Such technique is in use at NSLS-II [7]. This note describes the principle of the measurements and some details of how they were implemented at PIP2IT.

## 2. Procedure for a single corrector

In a beam line or linear accelerator with uncoupled linear optics, the change of the beam position $\Delta x_j$ recorded by a BPM #$j$ at the longitudinal location $z_j$ is proportional to deflection by the dipole corrector $\Delta \theta_1$

$$\Delta x_j(z_j, t) = \Delta \theta_1 \sqrt{\beta_x(z_j)\beta_{x1}} \sin \varphi_j, \qquad (1)$$

where $\beta_{x1}$ and $\beta_x(z_j)$ are Twiss beta-functions in the location of the corrector and BPM, correspondingly, and $\varphi_j$ is the betatron phase advance between the corrector and the BPM.

Let the corrector current change with time in sinusoidal manner with frequency $fc = \frac{\omega}{2\pi}$ producing the deflection amplitude $\theta_1$, while recording $N_p$ BPM readings with frequency $fb$ at moments $t_k = \frac{k}{fb}$. The set of readings is

$$x_{jk} \equiv \Delta x_j(z_j, t_k) = a0_j \sin \omega t_k = \theta_1 \sqrt{\beta_x(z_j)\beta_{x1}} \sin \varphi_j \sin 2\pi k \frac{fc}{fb}. \qquad (2)$$

Application of the Discrete Fourier Transform (DFT) to the data set $\{x_{jk}\}$ delivers the frequency components $c_{jm}$ (as defined in MathCad):

$$c0_{jm} = \frac{1}{\sqrt{N_p}} \sum_{k=0}^{N_p-1} x_{jk} e^{i\frac{2\pi m}{N_p}k} \tag{3}$$

The oscillation frequency and the total number of sampling points can be chosen so that

$$N_p \frac{fc}{fb} = L, \tag{4}$$

where $L$ is an integer indicating the number of full oscillations periods during the measurement. In this case, direct substitution of Eq. (2) and Eq.(4) into Eq.(3) yields only two lines at the driving and aliasing frequencies:

$$c0_{jm} = \begin{cases} \frac{\sqrt{N_p}}{2} a0_j, & m = L, N_p - L \\ 0, & m \neq L, N_p - L, \end{cases} \tag{5}$$

The set of the measured oscillation amplitudes $\{a0_j\}$ for the given deflection defines the differential trajectory to be compared with simulations.

## 3. Error estimation

With a noise, the error of measuring the oscillation amplitude is determined by the strength of the noise component at the driving frequency. It can be estimated in approximation of a small white noise:
- The amplitude of all noise frequency components $a_{wn}$ have the same rms value for all $m$;
- The phase $\varphi_{wn}$ of each noise component is randomly distributed between -π and π;
- The noise amplitude is small in comparison with meaningful amplitudes, $a_{wn} \ll a0_j$.

In this case, the amplitude of a measured component is

$$a_j = |a0_j + a_{wn} \cdot e^{i\varphi_{wn}}| = \sqrt{a0_j^2 + a_{wn}^2 + 2a0_j a_{wn} \cos \varphi_{wn}}. \tag{6}$$

The resulting rms amplitude error $\sigma_a$ is calculated by averaging over multiple data sets:

$$\delta_{a0j}^2 = \overline{a_j^2} - \overline{a_j}^2 = \overline{a0_j^2 + a_{wn}^2 + 2a0_j a_{wn} \cos \varphi_{wn}} - \overline{\sqrt{a0_j^2 + a_{wn}^2 + 2a0_j a_{wn} \cos \varphi_{wn}}}^2 \approx$$

$$\approx a0_j^2 + \overline{a1^2} - \overline{\left(a0_j + \frac{a_{wn}^2}{2a0_j} + a1 \cos \varphi_{wn} - \frac{a0_{1j}}{8}\left(2\frac{a1}{a0_j}\cos \varphi_{wn}\right)^2\right)^2} =$$

$$= a0_j^2 + \overline{a1^2} - \overline{\left(a0_j + \frac{\overline{a_{wn}^2}}{2a0_j} - \frac{\overline{a1^2}}{4a0_j}\right)^2} \approx \frac{\overline{a_{wn}^2}}{2}. \tag{7}$$

Assuming that averaging over multiple sets and over the components in one set (excluding the driving and aliasing lines) gives the same result (i.e. the white noise), Eq.(7) estimates the rms error from a single measurement. It is the rms value of the not-driven Fourier components in the set divided by $\sqrt{2}$. Expressing the answer in terms of raw DFT components $c_{jm}$, the rms error is given by Eq. (8):

$$\delta_{a0j} = \sqrt{\frac{2}{N_p(N_p-2)} \sum_{m \neq L, N_p - L} c_{jm}^2}. \tag{8}$$

## 4. Emittance dilution

The slow drifts do not affect the results of oscillating trajectory measurements if their frequency is far from the driving one. Therefore, the signal-to-noise ratio (S/N) improves with the number of recorded points as $1/\sqrt{N_p}$, and the differential trajectory can be measured at small perturbations if the measurement is long enough. This gives an opportunity to perform such measurements in parallel with regular operation. Applicability of such scenario can be estimated assuming that the figure of merit is the effective (projected) rms emittance $\varepsilon_{eff}$ composed by averaging over the ensemble and time. In the location of the corrector, the effective emittance can be written as follows:

$$\varepsilon_{eff}^2 = \overline{x^2 \cdot (x' + \Delta\theta_1)^2} - \overline{x(x' + \Delta\theta_1)}^2 = \overline{x^2} \cdot \overline{x'^2} - \overline{xx'}^2 + \overline{x^2} \cdot \overline{\Delta\theta_1^2} \tag{9}$$

Averaging in Eq.(9) is made over the ensemble and time. Terms containing $\overline{x' \cdot \theta_1}$ are zero because oscillation and internal motion of the beam particles are uncorrelated, and averages of $x, x'$, and $\Delta\theta_1$ are zero. Denoting the rms emittance without oscillations as $\varepsilon_0 = \overline{x^2} \cdot \overline{x'^2} - \overline{xx'}^2$ and using that $\overline{\Delta\theta_1^2} = \frac{\theta_1^2}{2}$, Eq.(9) becomes

$$\varepsilon_{eff}^2 = \varepsilon_0^2 + \frac{\theta_1^2}{2} \cdot \overline{x^2}. \tag{10}$$

Using notation of Eq.(1) and Eq. (2), Eq. (10) can be transformed to measurable values of oscillation amplitude $a0_j$ and the rms beam size $\sigma_{xj}$:

$$\varepsilon_{eff}^2 = \varepsilon_0^2 \left(1 + \frac{a0_j^2}{2\sigma_{xj}^2 \sin^2 \varphi_j}\right). \tag{11}$$

For the case of small amplitude perturbations, the relative emittance dilution is

$$\frac{\Delta\varepsilon}{\varepsilon_0} \equiv \frac{\varepsilon_{eff}}{\varepsilon_0} - 1 \approx \frac{a0_j^2}{4\sigma_{xj}^2 \sin^2 \varphi_j}, \tag{12}$$

For a lattice without dramatic changes of the β-function, as it was in the PIP2IT case, the maximum oscillation amplitude is recorded around $\varphi_j \sim \pi(n + 0.5)$, and Eq.(12) with $\sin \varphi_j = 1$ gives a reasonable estimate. Therefore, to preserve the emittance, one needs to keep the measured maximum oscillation amplitudes much lower than the typical beam size. For a rough estimation, the emittance dilution is at 1% level if the beam oscillates by 10% of its rms size.

## 5. Independent excitation of multiple correctors

In a linear system, oscillation of a corrector with a pure sinusoid produces a single DFT spectrum line (ignoring the higher-frequency part of the Fourier spectrum with repeating aliasing lines) if condition of Eq.(4) is fulfilled. Ideally, one can oscillate simultaneously multiple correctors at different frequencies as soon as the number of periods is an integer less than half of measured points. Some of practical considerations limiting their maximum number and the frequency choice are as follows.

- Capability of the control system to synchronously set and read the devices may place a limit for the highest frequency of collecting the data points.
- The cheapest way to excite the trajectory oscillations is to use the regular correctors designed for DC use, which limits the maximum oscillation frequency to ~1 Hz.
- To decrease the signal-to-noise ratio (S/N), it's preferable to have multiple oscillations over the duration of the measurement, which puts the low-frequency limit.
- Some frequencies might be less preferential because of an increased noise at that specific frequency.
- The actual deflection may differ from a pure sinusoid. If so, it creates response at other frequencies, which may increase the noise for other excitations. A similar effect can be produced by propagation through non-linear beamline components. Therefore, the larger number of excitations might correspond to larger measurement errors. Note, however, that in most of measurements at PIP2IT it was not the case.
- One may prefer to have the driving frequencies separated to use the spectrum lines in between for measuring the noise level.
- Contributions from different excitations to dilution of the effective emittance are summed in quadratures. If the measurement is performed in parallel with regular operation so the beam quality needs to be maintained, the excitation amplitudes need to be decreased roughly as the number of excitation parameters in power ¼. To preserve the same S/N, the number of measured points need to be increased inversely proportional to square of the excitation amplitude. Hence, in this simplistic model the measurement time increases as a square root of the number of excitation parameters.

## 6. Direct measurement of optical functions

If optics in location of two dipole correctors in the same plane is known and stable, it can be used to visualize the Twiss beta-functions and phase advances corresponding to dipole motion downstream of the correctors. Let's considers a beam line (Fig. 1), where two correctors are separated by the betatron phase advance $\varphi_x$ and have known beta-functions $\beta_{x1}$ and $\beta_{x2}$. The correctors are oscillated at the same frequency $f = \frac{\omega}{2\pi}$ with the time phase shift $\varphi_t$ between them and deflection amplitudes $\theta_1, \theta_2$.

The BPM position reading in the location $z$ separated by the betatron phase advance $\varphi(z)$ and with beta-function $\beta_x(z)$ is the linear combination of responses to these two excitations:

$$x(z,t) = \theta_1 \sqrt{\beta_x(z)\beta_{x1}} \sin \varphi(z) \sin \omega t + + \theta_2 \sqrt{\beta_x(z)\beta_{x2}} \sin(\varphi(z) + \varphi_x) \sin(\omega t + \varphi_t). \quad (13)$$

As it is shown in Ref. [8], at the specific choice of the deflection amplitudes and the phase shift in time,

$$\theta_2 \sqrt{\beta_{x2}} = \theta_1 \sqrt{\beta_{x1}}, \quad \varphi_t = \pi - \varphi_x, \quad (14)$$

Eq.(13) is simplified to

$$x(z,t) = \theta_1 \sqrt{\beta_x(z)\beta_{x1}} \sin \varphi_x \sin(\omega t + \varphi_1(z)). \quad (15)$$

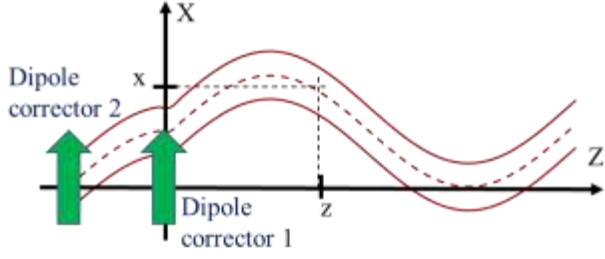

Figure 1. Model with two oscillating dipole correctors.

The squared value of the Fourier component amplitude in the BPM position is proportional to the zero-current beta-function in the location of the BPM, and its phase is equal to the betatron phase advance (with $2\pi n$ uncertainty).

If the effect of space charge between the excitation and observation is negligible, the beta-functions of the beam centroid (dipole) motion and of its envelope are the same. In this case, the choice of Eq.(14) means that the beam centroid is moved in time over an ellipse in the phase space that coincides, within a constant multiplier, with the equation for the rms particle:

$$\gamma_x x^2 + 2\alpha_x x x' + \beta_x x'^2 = \theta_1^2 \beta_{x1} \sin^2 \varphi_x . \tag{16}$$

Therefore, the amplitude of oscillations $a0$ is proportional to the beam rms size $\sigma_x$ at the same location

$$\sigma_x = \sqrt{\varepsilon_x \beta_x(z)} = \kappa \cdot a0; \; \kappa \equiv \frac{\sqrt{\varepsilon_x}}{\theta_1 \sqrt{\beta_{x1}} \sin \varphi_x} \tag{14}$$

with the coefficient of proportionality $\kappa$ being the same for all locations. Plotting in the control room the signals of BPM position amplitudes with such coefficient would give a live representation of the rms size along the beam line, which might be useful for tuning.

The procedure was tested at Fermilab Linac [9] and in the 2020/21 run of PIP2IT [10] with the main goal to use it for finding the loss locations. It appears to reproduce well the dipole motion but not the beam envelope, likely because of a large space chare at low energy.

Recently, Ref. [11] suggested using the relationships of Eq.(14) for optics measurements in the rings with the same idea of moving the beam around the phase ellipse.

## 7. Notes

### *7.1 Note on nonlinear elements*

Eq.(1) is valid if focusing elements of the beamline are linear. If it is not the case (for example, the base trajectory is significantly shifted and goes through a nonlinear area of a magnet), positions recorded by BPMs are not proportional anymore to the initial deflection. As a result, the Fourier spectrum contains, in addition to the main frequency, its harmonics. The longitudinal position, where such component appears, and the component's amplitude point to the location of the nonlinear element and strength of the non-linearity.

*7.2 Note on longitudinal measurements*

For a non-relativistic ion beam, similar measurements can be performed in the longitudinal plane. The analog of a transverse deflection is an energy change resulting from oscillating either phase or field amplitude of a cavity. The positions are replaced by the bunch phases reported by BPMs. Amplitudes of their Fourier components present the differential trajectory in the longitudinal plane and can be compared with simulations of longitudinal dynamics.

*7.3 Note on losses*

During these measurements, it is useful to record and analyze the BPM sum signals as well. If the beam is partially lost between the points of excitation and position measurement, a component at the driving frequency appears in the sum signals of all BPM downstream of the loss location (similar to observations in Ref. [8]). It would be a warning sign, since a large loss means that different BPMs measure the beam with different characteristics. Also, if beam or secondary particles can reach the BPM plates, their current may modify the normally purely capacitive signal, and BPM position readings might do not correctly present the position of the beam centroid. Therefore, measurements where the BPM sums do not exhibit a large signal at the driving frequency are preferable.

*7.4 Note on periodicity in time*

The procedure does not depend explicitly on how the corrector current changes with time. Only synchrony between corrector changes and data recording is required, and the Fourier transform is performed with respect to the points where the corrector current was changed. It mattered at the PIP2IT, where the correctors and BPMs front ends were not well synchronized. The control system acquired data only after acknowledgement that the corrector current references have been changed, and the time intervals between acquisition points could vary. Nevertheless, the procedure still worked for the employed slow variations.

However, the constant time intervals are generally preferable. For higher frequencies, the beam deflection can deviate from the reference signal because of the power supply bandwidth, corrector inductance, and vacuum chamber skin effect. If settings are performed sinusoidally in time, the deflecting magnetic field still changes mainly at the driving frequency, only with a decreased amplitude and a delayed phase. What usually matters most is the relationship between responses in different BPMs, which is not affected, and corresponding corrections to the corrector calibrations can be applied.

## 8. Examples

Measurements of this type were routinely used at the PIP2IT during 2020/21 run. Sets with periodically changing parameters were prepared in advance as a csv file, and then a Java program [12] was setting them, recording the specified signals after each set point was confirmed. The results, saved as a csv file, were analyzed offline.

A typical measurement was to oscillate one or two dipole correctors in the warm Medium Energy Beam transport (MEBT) section of the accelerator and record responses in all available BPMs. In contrast, two examples below were dedicated tests that may show better the specific realization and capabilities of the method.

*8.1 Oscillating of 6 parameters*

While typically only one or two parameters of interest were changed in a one measurement, a test with exciting of 6 parameters at once was performed as well. Data from this test are illustrated by Fig. 2. Data analysis was performed in MathCad.

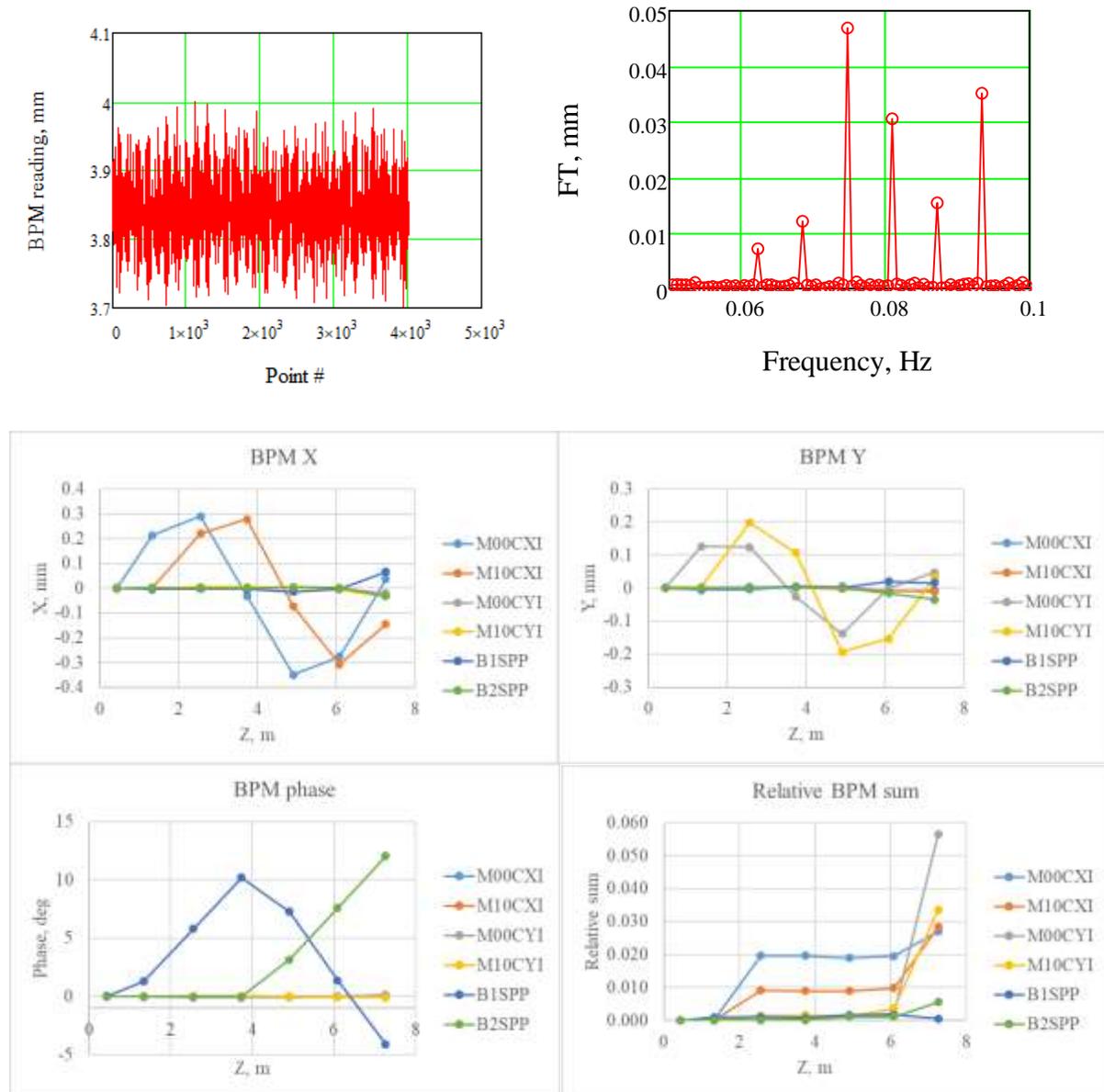

Figure 2. Response of BPMs to simultaneous excitation of two horizontal correctors (M00CXI and M10CXI), two vertical correctors (M00CYI and M10CYI), and two bunching cavities (B1SPP and B2SPP). See the text for description.

In the test, currents of 4 dipole correctors and phases of two bunching cavities in PIP2IT MEBT were oscillated at different frequencies. The parameters were changed in 10 consecutive identical cycles of 401 points, for the total of 4010 points. Parameters were changed with 6 different periods per cycle: 10, 11, 12, 13, 14, and 15. Signals were recorded from 7 BPMs, each reporting X and Y position, bunch phase, and sum signals of the 1st and 3rd harmonics. BPMs were reporting the data to the control system at the beam pulse frequency of 20 Hz, but only points corresponding to the confirmed settings were saved in the program's output file. As a result, the average setting and recording rate was about 2.5 Hz, and the total measurements time was 27 min. The measurements were performed with the 2.1 MeV H- beam at the pulse length of 10 µs. Typical rms beam size in the MEBT was 2 mm.

The top row of Fig. 2 illustrates the raw vertical signal of the most downstream vertical BPM (left) and a part of its Fourier spectrum (right). The Fourier components (vertical scale) are normalized as $a0_j$ in Eq.(5) to represent the amplitudes, in mm, of oscillatory motion at the given frequency.

The middle row shows horizontal and vertical BPM amplitudes along the beam line. The sign is determined by the sign of the Fourier component phase. The horizontal axis represents positions of the BPMs counted from the beginning of the MEBT.

The bottom left picture shows the BPM bunch phase.

The bottom right plot is the relative sum signal of the first harmonic (the ratio of the Fourier component to its average value). The locations of signal jumps correspond to known aperture limitations in corresponding planes.

Statistical errors in the plots, estimated similar to Eq. (8) (but subtracting now all 6 driving frequencies), are lower than the markers size. They range 0.1 – 0.7 µm for the positions and 0.002 – 0.02 degree for phases. The values change inversely with the square root of the number of cycles considered in analysis (Fig. 3), in agreement with observation that the scatter comes from the white (in this part of spectrum) noise produced by the beam jitter and BPM electronics.

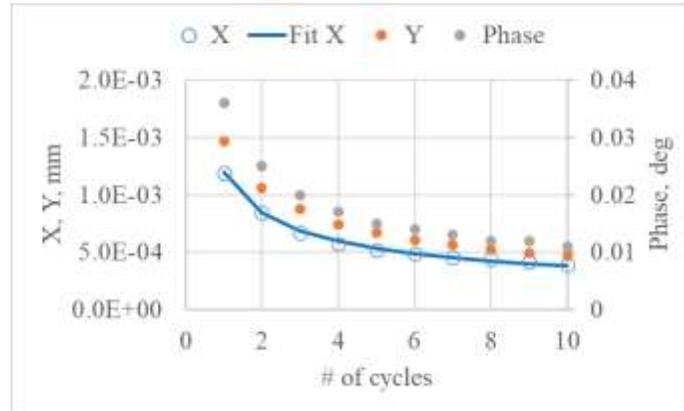

Figure 3. Dependence of rms errors on the number of cycles included into the analysis. The markers show the rms error values averaged over all BPMs for X, Y, and phase channels. The solid curve is the $1/\sqrt{N_p}$ fit for the X data.

Note that most of regular measurements were performed with a single cycle and lower amplitudes of oscillations since µm- type errors were satisfactory.

It appears that the total number of driving signals could be increased significantly. Extending the band to higher frequencies by a factor of 10 (to ~1 Hz) doesn't contradict to any limitations. The driving lines in Fig. 2 are separated by 9 empty lines, and exciting 4 of them (to still have empty lines in between to monitor the noise) would increase the number by 5. In total, simultaneous oscillation of up to 300 lines sounds plausible.

*8.2 Oscillating with a distorted signal*

Almost all measurements with oscillating trajectories were made using components in the warm front end of the PIP2IT. Fig. 4 shows the example when oscillations were applied to a superconductive dipole corrector inside a cryomodule.

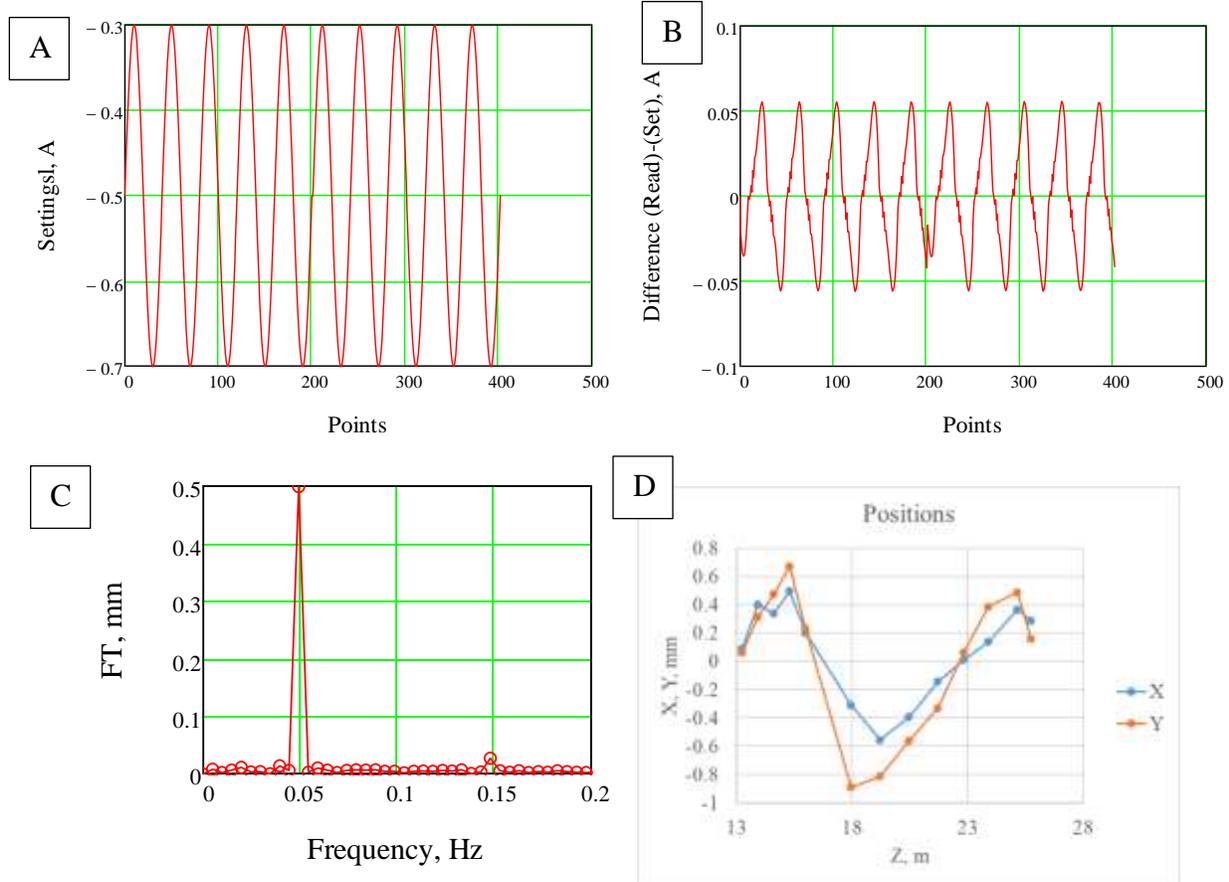

Figure 4. Differential trajectory measurement with a superconducting dipole corrector (HW4CX). 404 total points; recording frequency is 2 Hz. A- settings, B- difference between the current of the corrector and its settings, C – relevant part of the Fourier spectrum of a BPM (HR7BX, Z= 15.3 m), D – reconstructed differential trajectory. The rms errors are 1 – 6 µm.

A significant difference with warm correctors was that the cold correctors system was design to avoid fast changes in their currents to prevent quenching. Likely because of that, readings of the actual corrector current differ noticeably from settings (unlike the case of the warm correctors). In the example shown in Fig. 4, the amplitude of the actual current oscillation is 5% lower, its $1^{st}$ harmonic signal is shifted by 0.23 rad from the settings, and the $3^{rd}$ harmonic is clearly visible in the current signal as well as in the BPMs. Nevertheless, the BPM readings at the $1^{st}$ harmonic have a good signal-to-noise ratio, errors in the reconstructed differential trajectory are low, and such measurements can be used for optics analysis. Obviously, for accurate calibration of the corrector deflection, one needs to use readings of the corrector current rather than settings.

## 9. Conclusion

Measuring the optics of a linac with oscillating trajectories appears to be an effective and accurate way to find imperfections. A possible consistent scheme is as follows:
- Use regular "slow" dipole correctors for trajectory excitations at low frequency
- Keep the trajectory oscillation amplitudes low, so perturbations from such measurements are low enough to do not affect a routine operation.
- In this case, the measurement time can be long enough to provide a good measurement accuracy.

The measurements can be performed with oscillating multiple correctors at the same time at different frequencies. With thousands of measured points, it might be possible to make hundreds of measurements as the same time.

In an ion linac, the technique is applicable for measuring the longitudinal dipole dynamics as well.

## 10. Acknowledgement

## 11. References

1. W.J. Corbett, M.J. Lee and V. Ziemann, A fast model calibration procedure for storage rings, Proc. of PAC'93, p. 108
2. Accelerator Physics at Tevatron Collider, edited by V. Lebedev and V. Shiltsev, Springer, 2014
3. Experimental determination of storage ring optics using orbit response measurements, J. Safranek, BNL, NIM-A388, (1997) 27-36
4. R. Tomás, M. Aiba, A. Franchi, and U. Iriso, Review of linear optics measurement and correction for charged particle accelerators, Phys. Rev. Accel. Beams 20, 054801 (2017).
5. Linear optics correction for linacs and free electron lasers, Tong Zhang, Xiaobiao Huang, and Tim Maxwell, SLAC, PR-AB 21, 092801 (2018)
6. P. Derwent et al., PIP-II Injector Test: Challenges and Status, in Proc. of LINAC'16, USA, 2016, p. 641, WE1A01
7. Fast and precise technique for magnet lattice correction via sine-wave excitation of fast correctors, X. Yang, V. Smaluk, L. H. Yu, Y. Tian, and K. Ha, Phys. Rev. Accel. Beams 20, 054001, (2017)
8. A. Shemyakin, Finding beam loss locations in a linac with oscillating dipole correctors, FERMILAB-TM-2708-AD (2019); http://arxiv.org/abs/1905.00363
9. A.V. Shemyakin, K. Seiya, R. Prakash, Finding Beam Loss Locations in a Linac with Oscillating Dipole Correctors, Proc. of NAPAC'19, East Lansing, MI, USA, paper WEPLM02

10. A. Shemyakin, Finding a loss location with oscillating trajectories at PIP2IT, internal report, PIP-II Document 548 (2021)
11. Linear optics and coupling correction with closed orbit modulation, Xiaobiao Huang, Phys. Rev. Accel. Beams 21, 072805 (2021)
12. The data acquisition program was written by W. Marsh